\documentclass[12pt,preprint]{aastex}
\usepackage{bm}
\newcommand{\eb}{\begin{equation}}
\newcommand{\ee}{\end{equation}}

\newcommand{\kms}   {\ensuremath{ \mbox{km\,s}^{-1}                 }}

\newcommand{\kmspc} {\ensuremath{ \mbox{km\,s}^{-1}\,\mbox{pc}^{-1} }}

\newcommand{\masyr} {\ensuremath{ \mbox{ $\mu$as\,yr}^{-1} }}

\newcommand{\uasyr} {\mbox{$\mu$as~yr$^{-1}$}}
\newcommand{\uas} {\mbox{$\mu$as}}
\shorttitle{Secular aberration}
\shortauthors{Kopeikin \& Makarov}

\begin{document}

%%\center{Draft version \today, \  To be submitted to AJ Letters}
 
\title{Astrometric Effects of Secular Aberration}

\author{Sergei M. Kopeikin \altaffilmark{1} \& Valeri V. Makarov  \altaffilmark{2}}
\affil{$^1$Department of Physics and Astronomy, University of Missouri-Columbia,
Columbia, Missouri 65211}
\affil{$^2$Michelson Science Center, California Technology Institute, 770 S. Wilson Ave.,
MS 100-22, Pasadena, CA 91125}
\email{kopeikins@missouri.edu, vvm@caltech.edu}

\begin{abstract}
One of the main endeavors of fundamental astrometry is to establish a practical realization of a non-rotating, inertial reference frame anchored to celestial objects whose positions are defined in the barycentric coordinates of the solar system matching the current level of astrometric observational accuracy. The development of astrometric facilities operating from space at a microarcsecond level of precision makes the non-uniformity of the galactic motion of the barycenter an observable and non-negligible effect that violates the 
desired inertiality of the barycentric frame of the solar system. Most of the observable effect is caused by the nearly-constant (secular) acceleration of the barycenter with respect to the center of the Galaxy. The acceleration results in a pattern of secular aberration which is observable astrometrically as a systematic vector field of the apparent proper motions of  distant quasars.

We employ the classic approximations of planar epicycle and vertical harmonic oscillation for the Sun's galactic motion to estimate the magnitude of secular acceleration components in the galactic coordinates and show that these 
approximations are adequate for the SIM space mission. We employ the vector spherical harmonic formalism to describe the predicted field of  proper motions and evaluate the amplitude of this field at each point on the celestial sphere. It is shown that the pattern of 
secular aberration is fully represented by three low-order electric-type vector harmonics, and hence, it is easily distinguishable from the residual rotations of the reference frame and other possible effects, such as the hypothetical long-period gravitational waves, which are described by other types of vector or tensor harmonics. Comprehensive numerical simulations of the grid astrometry with 
SIM PlanetQuest are conducted assuming that 110 optically bright quasars are included as grid objects and observed on the 
same schedule as regular grid stars. The full covariance matrix of the simulated grid solution is used to evaluate the covariances 
of the three electric harmonic coefficients, representing the secular aberration pattern of proper motions. We conclude that the grid astrometry with SIM PlanetQuest will be sensitive to the main galactocentric component of secular acceleration, arising from the circular motion of the Local Standard of Rest (LSR) around the galactic center, while the peculiar acceleration of the Sun with 
respect to LSR is expected to be too small to be detected with this astrometric space interferometer. 
\end{abstract}

\keywords{astrometry --- techniques: interferometric ---
 reference systems --- Galaxy: kinematics and dynamics --- quasars: general --- dark matter}

\section{Introduction}

The Sun accelerates as it travels in the galaxy. The acceleration is small
(approximately, $6$ mm s$^{-1}$ yr$^{-1}$),
and it takes millions of years to change the Sun's velocity vector significantly.
The instantaneous velocity of the Sun in a non-rotating galactic reference
frame with the origin at the center-of-mass of the Milky Way causes all observed directions to both galactic and extragalactic sources to deflect from their
``true" directions in a systematic way. This phenomenon has the same special-relativistic nature as the annual
stellar aberration induced by the orbital motion of the Earth around the Sun. Due to the aberration the observed 
position of a light source is displaced toward the direction of the instantaneous velocity of the observer with respect to an inertial reference frame at rest by an
amount proportional to the velocity magnitude \citep{ks}.

For an observer located at the barycenter of the solar system, the instantaneous effect of the relativistic aberration due to the galactic motion of the solar system is not directly
observable at any fixed instant of time because the velocity-induced aberration pattern is constant. 
But since the motion of the solar barycenter with respect to a
fixed, non-rotating galactic reference frame is non-rectilinear, the quasi-stationary
(secular) change of its velocity vector causes the aberration pattern to change on the sky gradually as time progresses.
Given astrometric measurements at a microarcsecond (\uas) level of accuracy conducted over a time span
of at least several years, the changing aberration pattern can be observed in two
ways. An astrometric facility, such as Very Long Baseline Interferometry, capable of measuring long arcs between
light sources at two separate epochs, could directly detect and measure
the distortion of the grid of a global astrometric frame based on quasars taken as reference objects \citep{freid}. This distortion would be a direct consequence of
the directional difference in Sun's galactic velocity between the two epochs, as well as other possible
phenomena, e.g., unconstrained residual rotations of the reference frame, gravitational lensing from invisible dark matter 
\citep{liddle}, long-periodic gravitational waves of cosmological origin \citep{pyne1,pyne2}, etc.

A more tenable approach to detecting the secular aberration is to observe the global vector field of the proper motions of quasars provoked by the slowly changing aberration pattern that is induced by the secular acceleration of the Sun (see Fig. \ref{pmfield.fig}). 
This idea has circulated among astrometric community for a number of years \citep{gaia-report}. Nowadays it becomes 
technologically 
feasible and can be rendered within the framework of the existing projects of space astrometry which are currently under development in the national \citep{sim} and European space agencies \citep{gaia}.

Proper motions of stars projected on the celestial sphere are customary expressed in angular units (e.g. arcseconds) per year, and 
stem from the spatial motion of stars 
with respect to the solar barycenter. Physical proper motion of each star is proportional to the tangential, that is orthogonal to the line of sight, component of the spatial velocity and inversely proportional to the distance to the star. For this reason, physical proper motions
of quasars, which are separated from the Sun by distances of many Megaparsecs, are orders of magnitude
smaller than 1 \uasyr, and, thus, will be hardly observable in the near future. The secular aberration, however, mimics the physical proper motion of a distant celestial object and
makes all of them, including ``infinitely" distant quasars, move on the sky with
observable proper motions of $\approx 4$ \uasyr, which can be detected with current technology.
Measuring the secular aberration is vitally important for evaluating the total mass of the Milky Way including both visible and invisible matter in its bulge and halo which will provide us with a precise estimate of the amount of dark matter in our Galaxy.  

This paper discusses the astrometric effects of secular aberration which can be detected by the SIM space mission. 
The basic relationship between the galactocentric acceleration of the solar system and the observed pattern of proper motions 
of distant quasars
is established in section \ref{pm.sec}. The planar epicycle theory and a
harmonic vertical-motion approximation are used in section \ref{ep.sec} to evaluate the peculiar acceleration
of the Sun with respect to the Local Standard of Rest (LSR). A decomposition of the resulting proper motion field in
orthogonal vector spherical harmonics is given in section \ref{vec.sec}.
The vector-harmonics technique and the simulated global covariance
matrix of a number of quasars potentially observed with the Space Interferometry Mission are used to compute the expected accuracy of measuring the secular acceleration for this project (section \ref{sim.sec}).

\section{Proper motions and secular aberration}
\label{pm.sec}

Let us choose barycentric Cartesian coordinates $(X,Y,Z)$ such that the plane of the galaxy is 
the $X-Y$ plane and the $X$ axis is directed towards the center of the galaxy. 
The axis $Y$ is aligned with the direction of the galactic rotation.
Let us introduce spherical coordinates $(r,l,b)$ such that the unit basis vectors of the two coordinate systems are 
related by the following equations \citep{binney}
\begin{eqnarray}\label{1}
\vec e_r&=&\cos b\cos l\, \vec e_X+\cos b\sin l\, \vec e_Y +\sin b\, \vec e_Z\;\\\label{} 
\vec e_l&=&-\sin l\, \vec e_X+\cos l\, \vec e_Y \;\\\label{} 
\vec e_b&=&-\sin b\cos l\, \vec e_X-\sin b\sin l\, \vec e_Y +\cos b\, \vec e_Z\;,\\\nonumber 
\end{eqnarray}
where the angular coordinates $(l,b)$ are the galactic longitude
and latitude respectively, and the direction to the galactic center is given 
by the unit vector $\vec e_X$.
The geometric direction to a star at coordinates $(r,l,b)$ is given by the 
unit vector $\vec e_r\equiv \vec K$. Let us assume that at a given epoch $t_0$ the orbital 
velocity of the Sun around the galactic center is $\vec{V}=(V_X,V_Y,V_Z)$ and the acceleration 
is $\vec{A}=(A_X,A_Y,A_Z)$.

Observed direction to the star measured by a moving observer differs from its true geometric position, $\vec K$, because of the stellar aberration that is proportional at each instant of time to the velocity of the observer with respect to the static frame. 
Since the velocity vector of the solar system is not constant due to the galactocentric acceleration, the aberration angle changes progressively. Therefore, at another epoch $t$, 
the star is seen by a fictitious observer located at the solar barycenter in the direction, $\vec k$, given by
\begin{equation}\label{2}
\vec k =\vec K+\frac{1}{c}\vec K\times(\vec V\times\vec K)+\frac{1}{c}\vec K\times(\vec A\times\vec K)\,(t-t_0)\;,
\end{equation}
where $``\times"$ denotes the Euclidean cross-product of two 3-vectors, $c$ is the speed of light in vacuum, 
$t_0$ is the initial epoch, $t$ is the time of observation, and we assume
that the contribution of the first and higher order derivatives of acceleration is negligible.
The first two terms in the right-hand side of Eq. (\ref{2}) are constant if one neglects the secular parallax \citep{binney}, 
which is irrelevant for quasars and will not be considered in this paper. The third term in the right-hand side of Eq. (\ref{2}) 
yields the secular aberration that makes all objects beyond the boundaries of the solar system to move in the sky with a 
proper motion 
\begin{equation}
\label{2a}
\vec\mu=\mu_{l}\,\vec e_l+\mu_b\,\vec e_b\;,
\end{equation}
where the longitudinal and latitudinal components of $\vec\mu$ are proportional to the acceleration of the solar system projected 
on the celestial plane
\begin{eqnarray}\label{3}
\mu_{l}&=&\frac{1}{c}\left(-A_X\sin l+A_Y\cos l\right)\;,\\\label{3a}
\mu_b&=&\frac{1}{c}\left(-A_X\sin b\cos l-A_Y\sin b\sin l+A_Z\cos b\right)\;.
\end{eqnarray}
If one assumes that the acceleration of the solar system barycenter has only $A_X=A$ component, the equations (\ref{2a})--(\ref{3a}) for the effective proper motion caused by the secular aberration are simplified and reduced to
\begin{equation}\label{4}
\vec\mu=-\frac{A}{c}\left(\sin l\,\vec e_l+\sin b\cos l\,\vec e_b\right)\;.
\end{equation}
Proper motion vectors $\vec\mu$ of a given number of objects represent a discrete vector field on the 
sphere which can be decomposed in a set of vector spherical harmonics \citep{thorne}. The largest
galactocentric component of the secular aberration can be determined from global astrometric observations as a systematic dipole component
of this vector field. For quasars the problem of its determination is simpler than for stars since they have negligibly small proper motions caused by 
their peculiar velocities with respect to the Hubble flow. Therefore, the secular aberration can be directly measured from the 
observed proper motions of quasars.
 
The magnitude of the secular aberration  effect in the case given by Eq. ({\ref{4}) is \citep{gaia-report}
\begin{equation}
\label{4a}
\mu=\frac{A}{c}\sin\zeta\;,
\end{equation} 
where $\sin\zeta=\sqrt{1-\cos^2l\cos^2b}$ and $\zeta$ is the angle between the direction towards the galactic center and that to the star. 
In what follows, we investigate a more
accurate approximation of Eq. (\ref{4}) that includes all three components of the Sun's acceleration,
and evaluate the effect of the secular aberration more adequately in terms of vector harmonics.

\section{Circular and peculiar accelerations of the Sun}
\label{ep.sec}

The velocity vector of the Sun in the galaxy is commonly considered to be the sum of two
components \citep{binney}. The first (and largest) component is the motion of the so-called, Local Standard of Rest
(LSR). By definition \citep{binney}, the LSR is involved in a circular planar motion around the center of mass of the galaxy at a constant rate with a period
$P_0$. The second component of the solar velocity is the differential (peculiar) motion of the Sun with respect to the LSR,
which makes the Sun's orbit to be non-circular and non-planar. We adopt a solar peculiar velocity of $\vec{V}_\sun=(10.0,5.3,7.2)$ \kms~ as given by \citep{deh}.
The absolute value of the rotational velocity of the LSR is
\eb
V_{\rm LSR}= 223\, \left( \frac{R_0}{8.5 {\rm kpc}} \right)\left( \frac{235 {\rm Myr}}{P_0}\right) \left[\kms\right]
\ee
where $R_0$ is the distance of the Sun from the galactic center and the basic quantities $P_0$ and $R_0$ are normalized to their best known values \citep{binney}. 

The galactocentric acceleration resulting from this circular motion is
\eb\label{ha}
a_{{\rm LSR}}=6.1\cdot10^{-6} \left( \frac{R_0}{8.5 {\rm kpc}} \right)\left( \frac{235 {\rm Myr}}{P_0}\right)^2
 \left[\kms~{\rm yr}^{-1}\right]\;,
\label{a_lsr.eq}
\ee
and the maximum value of a proper motion caused by this acceleration is
\eb\label{hb}
\mu_{\rm LSR}=4.2\, \left( \frac{R_0}{8.5 \,{\rm kpc}} \right)\left( \frac{235\, {\rm Myr}}{P_0}\right)^2\left[\masyr\right]\;.
\ee
As follows from equation (\ref{4a}) the maximum proper motion is achieved for objects lying on the great circle orthogonal to the direction to the galactic center. The proper motion vectors from this largest component
of the LSR acceleration are directed toward the galactic center (see Fig. \ref{pmfield.fig}).

In order to estimate the magnitude of the secular aberration caused by the peculiar acceleration of the solar system with respect
to the LSR, we employ the classic epicycle approximation for the planar
motion of the Sun, and a harmonic force approximation for its vertical oscillation
about the galactic plane \citep{marsuch,binney}. The peculiar velocity components in this model are \citep{mo}:
\begin{eqnarray}
\label{epi.eq}
\dot{X} & = & -X_0 \frac{kA}{B}\sin kt + \dot{X_0}\cos kt + \dot{Y_0}\frac{k}{2B}\sin kt \;,\\ \label{epi.eq2}
\dot{Y} & = & -X_0 \frac{2A(A-B)}{B}(1-\cos kt) - \dot{X_0}\frac{k}{2B}\sin kt + \dot{Y_0}\frac{1}{B}
(A-(A-B)\cos kt)\;, \\ \label{epi.eq3}
\dot{Z} & = & -Z_0 \nu\sin \nu t + \dot{Z_0}\cos \nu t\;, \\\nonumber
\end{eqnarray}
where the planar epicycle frequency $k=\sqrt{-4B(A-B)}$ is expressed in terms of the Oort's constants $A$ and $B$, $(X_0,Y_0,Z_0)$ are 
the galactic coordinates of the Sun with respect to the LSR at time $t_0$, and $(\dot X_0,\dot Y_0,\dot Z_0)$ are 
their velocity components.
The vertical frequency, $\nu=2\pi/P_\nu$, of the oscillatory motion of the solar system about the galactic plane is an ill-constrained parameter that depends on
the vertical period $P_\nu$, which we assume to be 60 Myr. The harmonic vertical oscillation
is a linearized approximation, which ignores the second-order terms in the expansion of the vertical component of the gravitational force 
over coordinate Z. According to
the current knowledge of the mass distribution in the Milky Way, this second-order terms are less than 3\% of the main 
harmonic term (which is proportional to Z)
within 70 pc of the plane \citep{mo}. Hence, they can be neglected in our estimations. The harmonic
acceleration derived from the equations (\ref{epi.eq})--(\ref{epi.eq3}) is accurate to about 0.05\% for the Sun, which is
lying at $9\pm 2$ pc above the galactic plane \citep{mars}.

The epicycle theory is also based on a linear approximation derived from the expansion of the local horizontal 
component of the gravitational force over radial distance from the LSR $R-R_0$. The non-linear
terms in the expansion of the horizontal force can reach 15\% of the magnitude of the linear term at a distances of 1 kpc from the LSR. The Sun does not stray farther than about 300 pc from the LSR in this direction. Our conjecture is that the quadratic terms in the radial-force expansion will be
less than 1.5\% of the epicyclic (linear) term. Because the quadratic terms correspond to the time
derivative of the solar peculiar acceleration, they appear to be negligibly small. Little (if anything)
is known about the non-radial local gravitational forces resulting from the influence of the galactic
bar, spiral arms and other possible massive attractors including dark matter. We do not attempt to include them
in  the calculations given in the present paper but a theoretical study of their impact on the peculiar components of the solar system motion would be highly desirable.

Differentiating equations (\ref{epi.eq})--(\ref{epi.eq3}) and setting $t=0$ and $X_0=0$  yields the solar peculiar acceleration
\begin{eqnarray}
\label{acc.eq}
a_X & = & -2(A-B) \dot{Y_0}\;, \\ \label{acc.eq2}
a_Y & = & 2(A-B) \dot{X_0}\;, \\ \label{acc.eq3}
a_Z & = & -Z_0 \nu^2 \;.
\end{eqnarray}
The corresponding numerical estimates of the peculiar acceleration components are, respectively
\begin{eqnarray}
\label{a_pec.eq}
a_X & = & -0.27\cdot 10^{-6}\left(\frac{(A-B)}{0.025\;\kmspc}\right) \left(\frac{\dot{Y_0}}{5.3 \;\kms}\right)\left[\kms~{\rm yr}^{-1}\right] \\ \label{a_pec.eq2}
a_Y & = & 0.51\cdot 10^{-6}\left(\frac{(A-B)}{0.025\;\kmspc}\right) \left(\frac{\dot{X_0}}{10.0 \;\kms}\right)\left[\kms~{\rm yr}^{-1}\right] \\ \label{a_pec.eq3}
a_Y & = & -0.11\cdot 10^{-6}\left(\frac{60 {\rm Myr}}{P_\nu}\right)^2 \left(\frac{Z_0}{10.0 \;{\rm pc}}\right)\left[\kms~{\rm yr}^{-1}\right]\;. 
\end{eqnarray}
Equations for the maximum proper motions corresponding to this peculiar acceleration expressed in \masyr, are obtained by
dividing equations (\ref{a_pec.eq})--(\ref{a_pec.eq3}) by  $1.45\cdot 10^{-6}$, which is a conversion factor of units used in the calculation. We find that the residual proper motion field caused by the peculiar acceleration of the solar system with respect to the LSR is smaller in amplitude than 
1 \masyr, that is about 10 times smaller than the secular aberration (\ref{hb}) caused by the galactocentric acceleration of the LSR shown in equation (\ref{ha}).

\section{Vector harmonic analysis of stellar proper motions}
\label{vec.sec}
A global discrete pattern of proper motions is a vector field $\vec\mu(l,b)$ on the celestial sphere which can be expanded 
in orthogonal vector functions of spherical coordinates   
\eb
\vec\mu(l,b)=\sum_{j=1}^\infty\left[t_j \vec{T_j}(l,b)+s_j\vec{S_j}(l,b)\right],
\ee
where $\vec{T_j}$ and $\vec{S_j}$ are orthogonal vector harmonics, called magnetic and
electric harmonics respectively \citep{thorne}, and $t_j$ and $s_j$ are the coefficients of the expansion. The vector harmonics $\vec{T_j}$ and $\vec{S_j}$ can be expressed in terms of the partial derivatives of the scalar
spherical harmonics $V_j$ \citep{arf} with respect to galactic longitude and latitude. Specifically, one has \citep{vit}
\begin{eqnarray}
\vec{T_j}(l,b)&=& C_j\left[ \frac{\partial V_j(l,b)}{\partial b}\vec{e_l}-\frac{1}{\cos b}\frac{\partial V_j(l,b)}{\partial l}\vec{e_b}\right]\label{op1}\\
\vec{S_j}(l,b)&=& C_j\left[ \frac{1}{\cos b}\frac{\partial V_j(l,b)}{\partial l}\vec{e_l}+\frac{\partial V_j(l,b)}{\partial b}\vec{e_b}\right]\label{op2}\;,
\end{eqnarray}
where $C_j\equiv C_{nm}$ are constant normalization coefficients making the harmonics orthonormal, and the cumulative index $j=\{nm\}$ counts the spherical harmonics $V_j\equiv V_{nm}$ over orders $n=0,1,\ldots,\infty$ and degrees $m=0,1,\ldots,n$. We do not specify the normalization constants $C_j$ because real observations will provide a vector field of proper motions 
sampled at a number of discrete points on the celestial sphere corresponding to the observed quasars. 
This constant is determined in the process of calculation of the covariance matrix of vector harmonic coefficients, as explained in 
the following text.

The scalar spherical harmonics are given in terms of the associate Legendre polynomials $P_n^m$ as follows  \citep{arf}
\begin{eqnarray}
V^c_{nm} &=& \sqrt{\frac{2n+1}{2\pi}\frac{(n-m)!}{(n+m)!}} P_n^m(\sin b)\cos ml\;, \label{}\\
V^s_{nm}    &=& \sqrt{\frac{2n+1}{2\pi}\frac{(n-m)!}{(n+m)!}} P_n^m(\sin b)\sin ml\;.
\end{eqnarray}
A useful identity for differentiating the Legendre
polynomials is
\eb\label{er}
\frac{\partial P_n^m(\sin b)}{\partial b}=\frac{1}{\cos b}[(n+1)\sin b P_n^m(\sin b)-(n-m+1)P_{n+1}^m(\sin b)]\;,
\ee
and it will be used in order to derive the vector spherical harmonics in equations (\ref{op1})--(\ref{op2}).

The first set of spherical harmonics generating a pair of vector harmonics, are the zonal harmonic $V_{10}\sim \sin b$ and the sectorial harmonics $V^c_{11}\sim \cos b\cos l$, $V^s_{11}\sim \sin b\sin l$. The vector harmonics corresponding to $V_{10}$ are
\begin{eqnarray}\label{qw}
\vec{T}_{10}(l,b)&=& C_{10}\cos b\;\vec{e_l}\\\label{qw1}
\vec{S}_{10}(l,b)&=& C_{10}\cos b\;\vec{e_b}\;.
\end{eqnarray}
The vector harmonics corresponding to $V^c_{11}$ are
\begin{eqnarray}\label{za}
\vec{T}_{11}^c&=& C_{11}(\cos l\sin b\;\vec{e_l}-\sin l \;\vec{e_b})\;, \\\label{za1}
\vec{S}_{11}^c&=& C_{11}(\cos l\sin b\;\vec{e_b}+\sin l \;\vec{e_l})\;,
\end{eqnarray}
and those corresponding to $V^s_{11}$ are
\begin{eqnarray}\label{ah}
\vec{T}_{11}^s&=& C_{11}(\sin l\sin b\;\vec{e_l}+\cos l \;\vec{e_b})\;,\\\label{ah1}
\vec{S}_{11}^s&=& C_{11}(-\cos l \;\vec{e_l}+\sin l\sin b\;\vec{e_b})\;.
\end{eqnarray}
Numerical factors $C_{10}$ and $C_{11}$ in Eqs. (\ref{qw})--(\ref{ah}) are constant normalization coefficients which are not required in our discussion.

The vector harmonic decomposition of the proper motion field defined by Eq. (\ref{2a}) can be given now in the following form
\eb\label{ko}
\vec \mu_a=\frac{1}{c}\left(-A_X\vec S_{11}^c-A_Y\vec S_{11}^s+A_Z\vec S_{10}^c\right)\;,
\ee
which elucidates that the proper motion field caused by the secular aberration is represented by the three low-order electric dipole harmonics of first order ($n=1$) and it is independent of the magnetic-type harmonics. This systematic vector field 
affects all observable directions and can be extracted from the randomly distributed vector field of stellar proper motions. However, this subtle effect is submerged
in the pattern of the physical proper motion field of galactic stars, which is roughly three orders of magnitude
larger than the secular aberration effect \citep{olde}. For quasars, because of their negligible physical proper motions, the systematic pattern of secular aberration
stands out clearly, and is a dominant component of the quasar proper motions. Fig. \ref{pmfield.fig} shows the sky distribution of the proper motion pattern caused by the acceleration of the solar system barycenter.

\section{Numerical Simulations of the Secular Aberration and Accuracy of Its Measurement by {\it SIM}}
\label{sim.sec}
\subsection{SIM Astrometric Grid Model} 
SIM PlanetQuest is a NASA Origins project dedicated mostly to the search of planets around
nearby stars by astrometric interferometry. As a necessary condition of achieving
this goal, SIM will create a global astrometric reference frame over all the sky to an unprecedented accuracy of
4-5 \uas~ in position, based on wide-angle measurements of about 1300 reference grid stars. It is crucial for the
astrometric coordinate grid to include a sufficient number of optically bright quasars, which constrain
the parallax solution of the astrophysical targets in such a way that the large-scale random distortions of the parallax error
distribution are drastically reduced \citep{mami}. The same grid quasars
can be directly used to determine the secular acceleration of the Sun from measuring the secular aberration effect as explained in the previous sections of this paper. 

To estimate the measurement accuracy of the secular aberration effect for SIM PlanetQuest we make use of
full-scale numerical simulations of SIM observations and the astrometric grid reduction model \citep{mami}. 
The simulation model is rather sophisticated 
and realistic in that it includes the SIM instrument model and incorporates the specific technology
of SIM astrometric measurements. The grid reduction model has about 160,000 free fitting parameters (unknowns)
and deals with nearly 315,000 numerically simulated delay measurements taken during supposed 5-year mission life span. 
The adopted one-step solution of the 
global astrometric grid is found by the least-square method which produces the
covariance matrix of fitting parameters. Among the original 160,000 parameters only 6,500 determine the astrometric grid, while the rest are field-dependent
instrument and baseline-orientation parameters \footnote{See \citep{mami} and
references therein.}. In this paper, we use only that part of the covariance matrix which contains 
astrometric information about the grid reference quasars directly related to determination of the secular aberration effect. 

Let us define a column array of vectors ${\bm\mu}=\{\vec\mu_1,\vec\mu_2,\ldots,\vec\mu_N\}$ comprised of the proper motion vectors 
of $N$ quasars.
We define a matrix  ${\bf S}$ representing $N\times 3$ array, whose columns are the vector harmonics taken at the points
$1,2,\ldots,N$. Let us also introduce a column vector $a\equiv\{a_i\}$ consisting of three coefficients $a_1$, $a_2$ and $a_3$ representing respectively $A_X,A_Y,A_Z$ components of the secular acceleration vector (see equation (\ref{ko})) which is to be determined.
These parameters are found in our model by the least squares method employed for solving the system of linear equations
\eb\label{ut}
{\bf S}\cdot a={\bm\mu}\;,
\label{ls.eq}
\ee
where the dot ``$\bm\cdot$" denotes the inner product with respect to the index $i=1,2,3$. If $\hat{a}$ is a least squares
solution of equation (\ref{ls.eq}), its covariance is \citep{bard}
\eb\label{cov}
{\rm Cov}[\hat a]={\rm E}\left[\hat a\hat a^T\right]=\left({\bf S}^T{\bf S}\right)^{-1}\left({\bf S}^T{\rm Cov}[{\bm\mu}]{\bf S}\right)\left({\bf S}^T{\bf S}\right)^{-1};,
\ee
where ${\rm E}[r]$ denotes the expectation value of random variable $r$, ${\bf S}^T$ is the transposed matrix 
${\bf S}$, and ${\rm Cov}[{\bm\mu}]={\rm E}[{\bm \mu}\cdot {\bm\mu}^T]$. We emphasize two points:
\begin{enumerate}
\item Solution $\hat{a}$ gives the components of the secular acceleration expressed in the same dimensional units as the observed 
proper motion vector field ${\bm\mu}$, that is in $\mu$as yr$^{-1}$. 
This choice of units determines coefficients $C_j$ in equations (\ref{op1})--(\ref{op2}) which are naturally fixed for a given discrete sample of points by the normalization
of the matrix ${\bf S}$ in equation (\ref{cov}). 
\item The vector spherical harmonics used for computing the fitting parameters $a_i$ are not orthogonal because they are sampled 
at a finite number of points on the celestial sphere. This accounts for the presence of terms ${\bf S}^T{\bf S}$ in equation (\ref{cov}). 
\end{enumerate}   

\subsection{Covariance Analysis}

Covariance ${\rm Cov}[{\bm\mu}]$ is a $N\times N$ matrix which is easily calculated from
the full covariance matrix of the astrometric grid solution \citep{mami} in the ecliptic coordinate system. 
In doing this, we do not have to rotate the covariance matrices from the ecliptic 
to the galactic coordinates because the inner products $(\bm\mu_k\cdot\bm\mu_m)$, for which the
covariance matrix in equation (\ref{cov}) is computed, are invariant with respect to rigid rotations of the coordinate system. However, the proper motion vectors should be projected
onto the galactic coordinate axes by making use of an appropriate
orthogonal transformation \citep[e.g.,][Vol. 1]{ESA97} before solving the system (\ref{ls.eq}), because
the vector harmonic functions are not invariant with respect to rotations of spherical coordinates.
We also notice that for any pair $(k,m)$ of the grid quasars the expected value of the inner product of their proper motions, 
${\rm Cov}[{\bm\mu}]_{km}$, is the sum of the two corresponding covariances of the coordinate proper motion components. 

Following this algorithm, we carry out a full-scale numerical simulation of the secular aberration problem,
using the grid solution method described in detail in \citep{mami}.  We assumed in this paper that
a set of $N=110$ quasars having a fairly uniform distribution on the sky, is included in the SIM reference grid. The astrometric frame
consists of a sufficiently dense grid of reference stars and is required to carry out astrometric reductions
of the science target data collected in the wide angle regime. The need for a reference grid arises from the fact that the space
interferometer will measure the optical path delay of the light-wave front from celestial sources with respect to
its own baseline vector, which has to be determined from the same
observations in a self-consistent manner \footnote{Similar problem exists in radio interferometry \citep{sovers}. Its solution also relies upon self-consistent construction of a fundamental reference frame.}. The majority of the SIM grid objects will be distant red
giant stars of $m_V\simeq 11$ magnitude, spectroscopically vetted as non-binary stars. However, it is imperative to include in the grid a number of optically bright quasars. Having parallaxes much smaller than 1 \uas, the quasars will serve to constrain
the global parallax solution significantly reducing the zero-point and other large-scale
parallax errors for scientific targets. Positions of the grid quasars determined with SIM will be used to remove
the residual distortions of the grid reference system based on the red giants, with respect to the extra-galactic reference system based on the quasars \citep{john}. 
We note here, that in order to make the SIM reference frame truly inertial to accuracy better than
$1~ \uasyr$, the residual spin present in the proper motion field and caused by the indefinite rotation of the grid solution, must 
be removed as well. Rigid rotations (spins) of the reference system
around the coordinate axes $X$, $Y$ and $Z$ are represented by the first three
magnetic-type vector harmonics of the observed proper motion field \citep{vit}. 
Hence, residual spins are orthogonal to the secular aberration effect and, for this reason, can be neatly separated from it in the vector harmonic space. This point can be effectively demonstrated by comparing the pattern of the proper motion field induced by the secular aberration and shown in Figure \ref{pmfield.fig}, against the patterns of the proper motion fields induced by residual spins of the 
global solution about $X,Y,Z$ axes shown in Figures \ref{pmfield-X}, \ref{pmfield-Y}, and \ref{pmfield-Z} respectively.

The covariance matrix, ${\rm Cov}[\hat a]$, of the secular acceleration components is given in Table~\ref{tab1}. The diagonal elements of the matrix are the unit-weight variances
of the corresponding $X,Y,Z$ components of the proper motion field (\ref{ko}). For example, the standard deviation of 
$a_1$ in units of \uasyr is estimated as follows
\eb\label{ft}
\sigma[a_1]=\sqrt{\rm Cov_{11}[\hat a]} \sigma_0=0.076\,\sigma_0,
\ee
and similarly for other coefficients,
where $\sigma_0$ is the standard error of a single measurement (the interferometer path delay),
expressed in \uas. For quasars as bright as $m_V=14$ magnitude, single measurement
errors of $15\div 20$ \uas~ could probably be achieved without spending too much integration time on 
these objects. Hence, the largest component of the proper motion field in $\vec S_{11}^c$,
which is proportional to the galactocentric component $A_X$ of the solar acceleration, will be
determined with an error, $\sigma(A_X)\simeq 1.5$ \uasyr as follows from equation (\ref{ft}). This accuracy is 
sufficient for a statistically robust detection of the acceleration at a signal-to-noise ratio of roughly 3.

\section{Discussion}

We have seen that, upon observing 110 optically bright quasars as grid objects, SIM PlanetQuest
will be able to detect the major galactocentric component of the acceleration of the Sun. At the same time, 
the anticipated accuracy of SIM is not high enough to detect 
the peculiar acceleration of the Sun with respect to the LSR, nor to refine the fundamental constants
$P_0$ and $R_0$. In order to do this a dedicated interferometric mission is required. 

Proper motions of the grid stars and the science target stars will also
include the secular acceleration components. 
These effects will be swamped by the
stronger patterns of differential galactic rotation, asymmetric drift, local and large-scale
dilation, disk warp, perturbations from the spiral arms and large streams of galactic stars, so that stellar
proper motions will probably be useless to study the subtle secular aberration effects. The latter, nonetheless,
have to be removed from the observed proper motions of stars to achieve the highest
accuracy and unbiased estimates in the galactic dynamics studies. By the same argument, the residual spins of the stellar proper motion field (which, in fact, may be rather large) can be reduced to $\approx 1$ \uasyr~ 
accuracy from quasar measurements.

Rigid rotations (spins) of SIM's reference frame are parts of the null space of a global astrometric solution \citep{lat,mami}, and,
therefore, can not be determined from observations with SIM. They are confined to three
low-order magnetic vector harmonics, while the vector field of proper motions caused by the secular acceleration is represented by the
three electric harmonics of zero and first order. SIM will establish an optical reference frame based on a number of reference stars. 
In order to make this frame as inertial as possible observation of
bright quasars are of utmost importance for making the residual rotations of this frame appropriately
small. It is important  to ensure that the grid quasars are uniformly distributed on the sky
and each of them is observed to roughly the same precision, to keep the correlations between the 
astrometric parameters as small as those
calculated in our numerical simulations (Table~\ref{tab1}). These requirements prevent the random errors from propagating
between the constituents of the vector harmonic space. Higher-order harmonics of the
quasar proper motion field will also be of a vast interest to examine since they will provide an
independent estimation of the {\it external} accuracy of SIM.

In principle, some additional large-scale pattern in the observed proper motions of grid quasars may be induced by the 
non-stationary effect of gravitational bending of light caused by large clumps of invisible (dark) matter in our Galaxy. Apparent positions of all quasars in the sky are shifted from their true positions due to this effect. Since the Sun is moving through the Galaxy the angle of the gravitational deflection of light will gradually change because of the change in the amount of the matter concentrated near the line of sight towards the quasar. The field of the spurious proper motions generated by this gravitational effect will look similar to that given by Eq. (\ref{ko}). In other words, the gravitational deflection of light can, in principle, interfere with the secular aberration effect. However, according to our estimates, the magnitude of this effect is much less than 1 $\mu$as and can be neglected. Nevertheless, more precise study of the gravitational lensing by clumps of dark matter in our Galaxy would be desirable (see, for example, \cite{ld}).

%\label{con.sec}

\acknowledgments
The research described in this paper was in part carried out at the Jet Propulsion 
Laboratory, California Institute of Technology, under a contract with the National 
Aeronautics and Space Administration. Grant support from the Eppley Foundation for Research and Research Council of the University of Missouri-Columbia is greatly appreciated.

\clearpage
\begin{figure}
\plotone{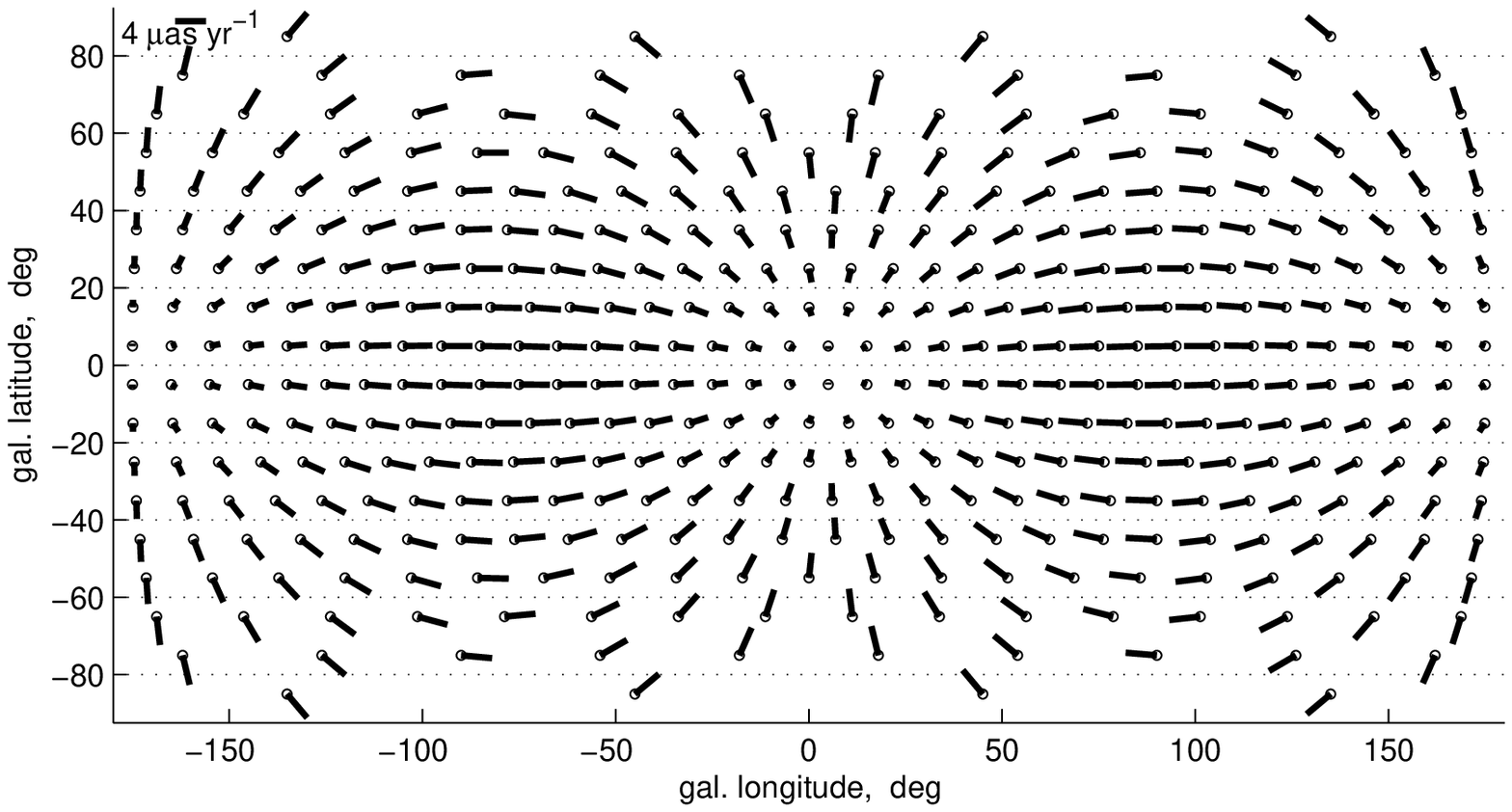}
\caption{Expected pattern of the proper motion field induced by the secular
acceleration of the solar barycenter. The length of a 4 $\mu$as/yr proper motion is shown in the
upper left corner.\label{pmfield.fig}}
\end{figure}
\clearpage
\begin{figure}
\plotone{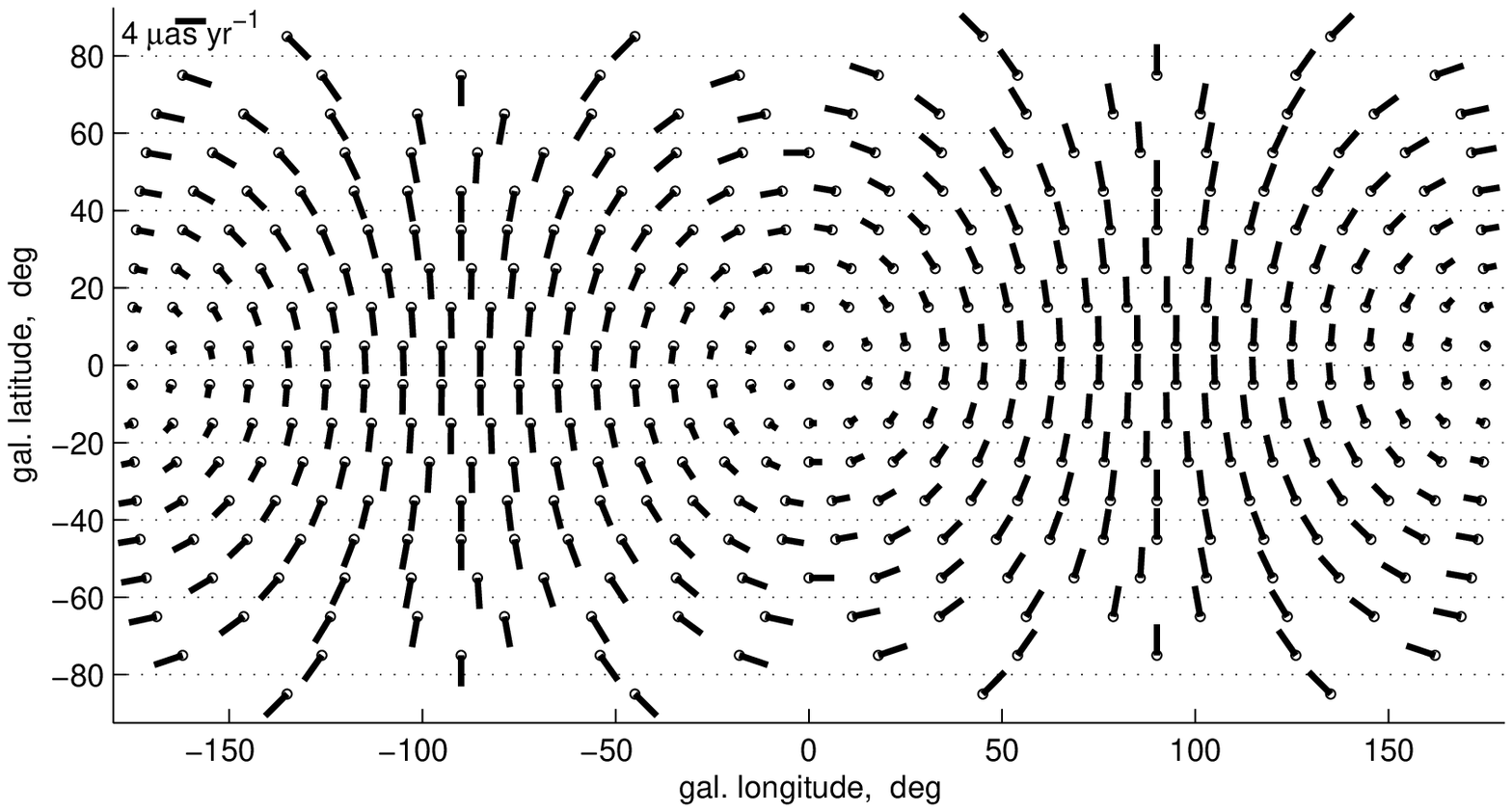}
\caption{Expected pattern of the proper motion field induced by the residual rotation of the coordinate system about X axis. The rotational rate is made comparable with the amplitude of the secular aberration.\label{pmfield-X} }
\end{figure}\clearpage
\begin{figure}
\plotone{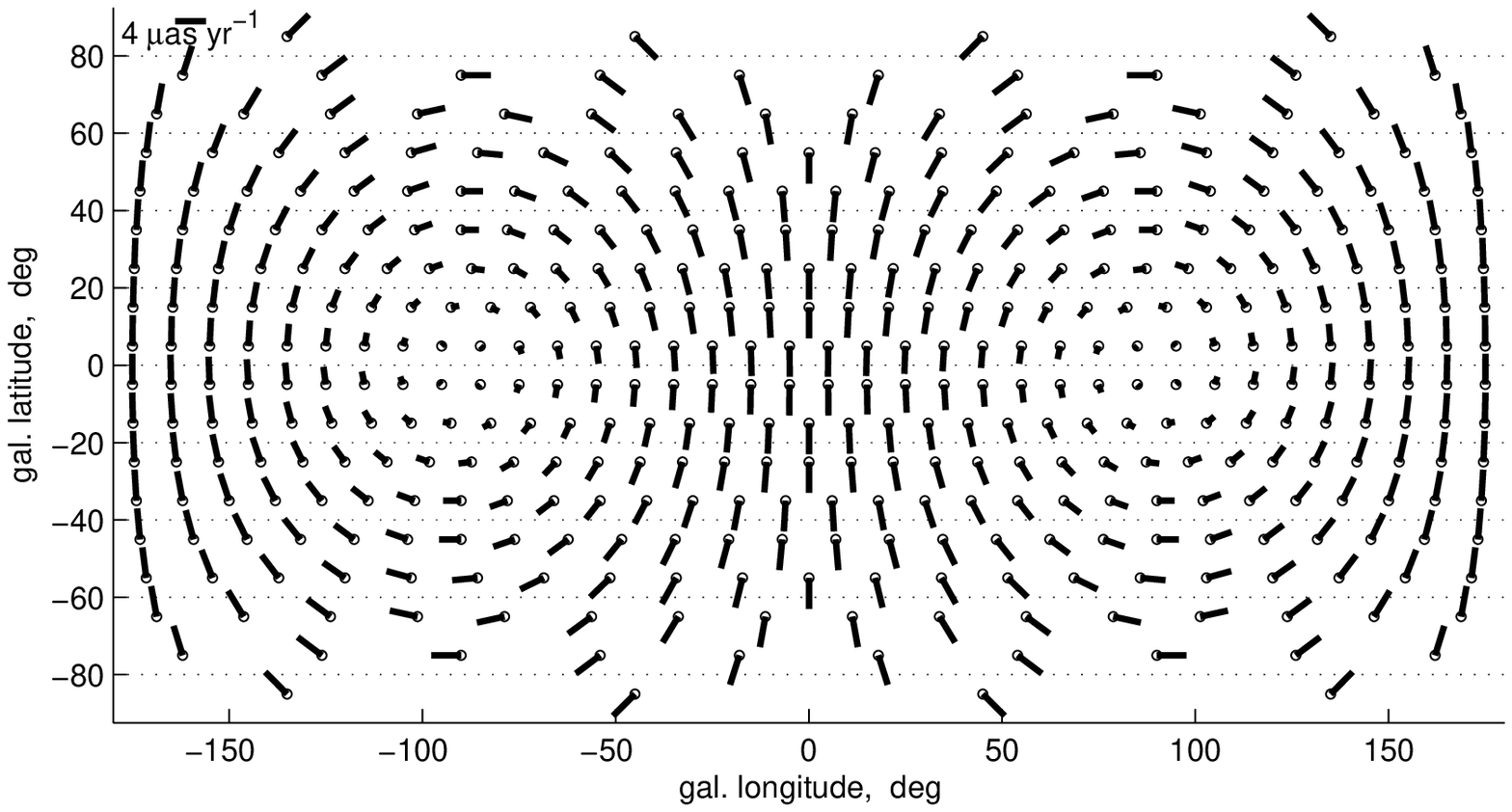}
\caption{Expected pattern of the proper motion field induced by the residual rotation of the coordinate system about Y axis. The rotational rate is made comparable with the amplitude of the secular aberration.\label{pmfield-Y} }
\end{figure}\clearpage
\begin{figure}
\plotone{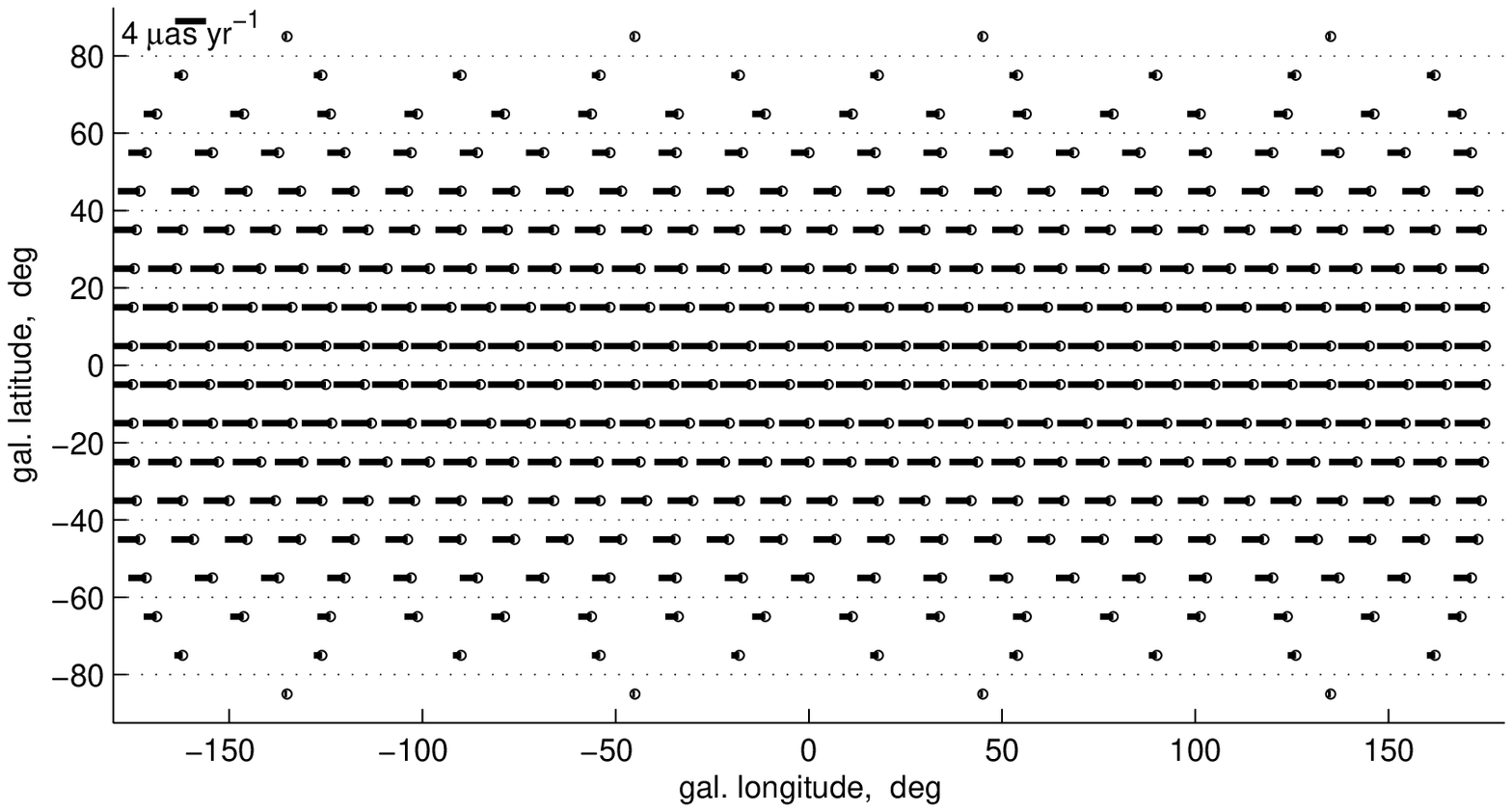}
\caption{Expected pattern of the proper motion field induced by the residual rotation of the coordinate system about Z axis. The rotational rate is made comparable with the amplitude of the secular aberration. \label{pmfield-Z}}
\end{figure}

\clearpage

\begin{deluxetable}{|l|rrr|}
\tabletypesize{\normalsize}
\tablecaption{Covariances of the secular acceleration components\label{tab1}}
\tablewidth{-0.2pt}
\tablehead{
 & \colhead{$a_1$} &  \colhead{$a_2$}  & \colhead{$a_3$}\vline }
\startdata
$a_1$ & 0.00578& $ -0.00050$ & $0.00003$  \\
$a_2$ & & $ 0.00412$ & $0.00000$  \\
$a_3$ &  & & $0.00199$  \\
\enddata
\end{deluxetable}
\end{document}